\documentclass{article}
\usepackage{spconf,amsmath,graphicx}

\usepackage{enumitem}
\setlist{nosep, leftmargin=14pt}

\usepackage{mwe} 
\usepackage{amssymb} 
\usepackage{bm} 
\usepackage{color}

\title{Disease Severity Regression with Continuous Data Augmentation}
\name{Shumpei Takezaki$^{1}$ \qquad Kiyohito Tanaka$^{2}$ \qquad Seiichi Uchida$^{1}$ \qquad  Takeaki Kadota$^{1}$\vspace{-2mm}}
\address{$^{1}$ Kyushu University, Fukuoka, Japan \\
    $^{2}$ Kyoto Second Red Cross Hospital, Kyoto, Japan\vspace{-2mm}}
 
\begin{document}
%
\maketitle
%
\begin{abstract}
Disease severity regression by a convolutional neural network (CNN) for medical images requires a sufficient number of image samples labeled with severity levels. Conditional generative adversarial network (cGAN)-based data augmentation (DA) is a possible solution, but it encounters two issues. The first issue is that existing cGANs cannot deal with real-valued severity levels as their conditions, and the second is that the severity of the generated images is not fully reliable. We propose {\em continuous DA} as a solution to the two issues. Our method uses {\em continuous severity GAN} to generate images at real-valued severity levels and {\em dataset-disjoint multi-objective optimization} to deal with the second issue. Our method was evaluated for estimating ulcerative colitis (UC) severity of endoscopic images and achieved higher classification performance than conventional DA methods.
\end{abstract}
\section{Introduction}
\label{sec:intro}
Disease severity regression is a task to determine a function $f(\bm{x})$ that satisfies $y_n \sim f(\bm{x}_n)$ for a given dataset $\Omega=\{(\bm{x}_n, y_n), n\in\{1,\ldots, N\}\}$, where $\bm{x}_n$ is a medical image, such as an endoscopic image, and $y_n$ is its severity level. Nowadays, it is common to use a convolutional neural network (CNN) as the model of $f(\bm{x})$ because CNN has a powerful representation ability to deal with the nonlinear relationship between image appearance and its severity. It is also common to use $L$ discrete severity levels as $y_n$. For example, Mayo scores of endoscopic images with ulcerative colitis (UC) have $L=4$ levels. 
\par
If the labeled dataset $\Omega$ is too small to train the CNN, data augmentation (DA) is often employed to generate synthetic data 
$\Omega’=\{(\bm{x}'_m, y'_m), m\in \{1,\ldots, M\}\}$ from $\Omega$. A possible DA technique is a conditional generative adversarial network (cGAN). Given a {\em discrete} severity level $y'_m \in \{1,\ldots,L\}$ as the condition, cGAN generates various images $\bm{x}'_m$ at the severity level $y'_m$. The generated images $\Omega'$ are then used to train the CNN together with the original dataset $\Omega$.\par
This paper focuses on two issues of the above cGAN-based DA for disease severity regression. The first issue is that disease severity is inherently continuous, so we do not need to adhere to the discrete conditions as $y'_m$. In other words, generating images at real-valued severity levels $y'_m$ will help train the CNN $f$ appropriately. The second issue is that the severity of the generated image is not very reliable. Even if we generate an image $\bm{x}'_m$ with the condition $y'_m$, there might be a risk that the visual severity of $\bm{x}'_m$ is precisely equal to $y'_m$.\par
\begin{figure*}[t]
    \centering
    \includegraphics[width=\linewidth]{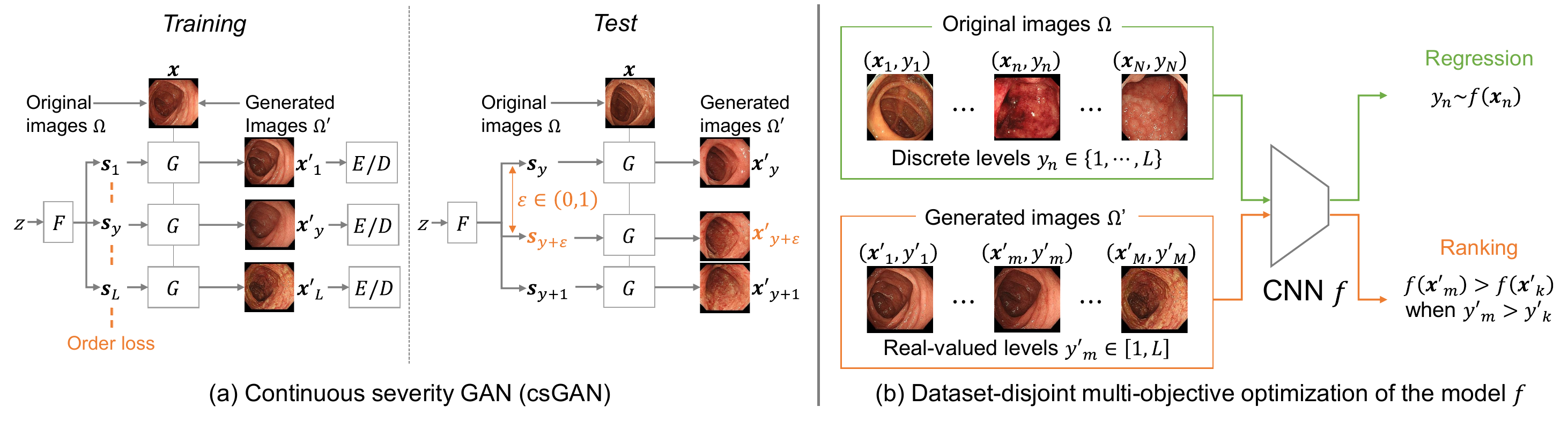}\\[-3mm]
    \caption{An overview of the two techniques of our continuous data augmentation scheme.}
    \label{fig:overview}
    \vspace{-3mm}
\end{figure*}

We propose a {\em continuous DA} scheme, where a new technique tackles each issue. For the first issue, we propose a {\em continuous severity GAN} (csGAN). Fig.~\ref{fig:overview}~(a) shows the overview of csGAN. Our csGAN is trained with images with discrete levels ($y_n \in \{1,\ldots, L\}$) but can generate images at real-valued severity levels ($y'_m \in [1, L]$). \par
For the second issue, we use a {\em dataset-disjoint} multi-objective optimization, where the original dataset $\Omega$ (with discrete levels) and the augmented dataset $\Omega'$ (with real-valued levels) are used in different ways according to their different reliability. Specifically, as shown in Fig.~\ref{fig:overview}~(b), we train a CNN $f$ with a regression loss for  $\Omega$ and a ranking loss for $\Omega'$. The former works to satisfy $y_n\sim f(\bm{x}_n)$ and the latter $f(\bm{x}'_m) \lessgtr f(\bm{x}'_k)$ when $y'_m \lessgtr y'_k$. This means that the levels $\{y'_m\}$ of the augmented data are not used as absolute ground truth but as relative conditions for training $f$.\par
The proposed techniques are evaluated by using a UC image dataset. As a qualitative evaluation, we observe the images by csGAN and confirm that we can continuously control the visual severity level of the generated images. As a quantitative evaluation, we confirm that our continuous DA helps to improve the severity regression performance.
\par
Our main contributions are summarized as follows:
\begin{itemize}
    \item We propose csGAN, which can generate images at real-valued severity levels.
    \item We also propose to use dataset-disjoint multi-objective optimization for the disease severity regression task with an augmented dataset.
    \item Experimental evaluations with a UC image dataset show the performance superiority of our continuous DA scheme using the above two techniques over a baseline and other cGAN-based DA.
\end{itemize}

\section{Related Work} \label{sec:rel_work}

\noindent{\bf Conditional GANs:}\ Various cGANs have been proposed so far, \cite{brock2018large,zhang2019self,odena2017conditional,gui2021} and they assume various types of conditions. For example, in the pix2pix~\cite{isola2017image} framework, an image is given as a condition. The most common condition is class labels -- they can be given as a one-hot vector or discrete number. In other words, for specifying the target type of generated images, it is not common to give a condition by a real-valued number (such as 1.33 and 0.28). Exceptionally, CcGAN~\cite{ding2020ccgan} accepts real-valued conditions; however, it relies on a hard assumption that real-valued annotation has already been attached to each training sample. In contrast, our csGAN can be trained with discrete conditions but still can generate images at real-valued conditions.\par
\noindent{\bf DA for medical images:} Due to a high cost for annotation, medical image analysis tasks often suffer from a limited number of labeled data and thus employ DA methods. According to a survey paper in 2021~\cite{chlap2021}, basic augmentation techniques, such as linear and nonlinear geometric transformations and intensity level perturbations, are still the majority for medical image DA. However, the survey also shows that GAN-based DA methods have increased in recent papers~(such as \cite{wang2020combination, ge2020enlarged, shin2018medical}). The above review for cGANs says that GAN-based DA for medical images has also not dealt with real-valued conditions. Moreover, to the authors' best knowledge, the augmented dataset is simply merged with the original dataset without any special treatment. 

\section{Continuous data augmentation} \label{ssec:cda}
This section describes two techniques in continuous DA, i.e., csGAN and dataset-disjoint multi-task optimization. The former is a new conditional GAN trained to generate images at real-valued severity levels. The latter is a technique to train the regression model $f$ by using the original dataset $\Omega$ and the augmented dataset $\Omega'$ in different manners by considering the reliability of the severity levels of the augmented data. 

\subsection{Continuous Severity GAN (csGAN)} \label{ssec:csgan}
As noted in Section~\ref{sec:intro}, we propose csGAN to generate images at real-valued severity levels. Inspired by StarGAN v2\cite{choi2020stargan}, csGAN comprises four modules: a mapping network $F$, a generator $G$, a style encoder $E$, and a discriminator $D$, as shown in Fig.~\ref{fig:overview}(a). csGAN uses a {\em style vector} $\bm{s}_y(\bm{z})$ as a condition to generate images at the level $y$, where $\bm{z}\sim \mathcal{N}(\bm{0},\bm{I})$. A different $\bm{z}$ results in a different style vector $\bm{s}_y(\bm{z})$ and finally contributes to having a different generated image at the level $y$. Hereafter, we often denote $\bm{s}_y(\bm{z})$ as $\bm{s}_y$ for simplicity.\par
Due to the page limitation,  we briefly summarize the roles of four modules $F$, $G$, $E$, and $D$:
\begin{itemize}
    \item $F$ accepts a random vector $\bm{z}$ and then outputs $L$ style vectors $\bm{s}_1, \ldots, \bm{s}_L$ at once. 
    \item $G$ accepts a real or generated (i.e., fake) image $\bm{x}$ and the condition $\bm{s}_y$ and then generates an image $\bm{x}'$ at $y$. 
    \item $E$ accepts a real or fake image  $\bm{x}$ with its level $y$ and then estimates its style vector while expecting the estimated vector is similar to $\bm{s}_y$ input to $G$. 
    \item $D$ is a standard discriminator for real/fake decisions of $\bm{x}$. 
\end{itemize}
Those modules are trained to achieve cycle consistency; a generated image $\bm{x}'=G(\bm{x}_n, \bm{s}_y)$ for the level $y\neq y_n$ needs to satisfy the condition $\bm{x}_n \sim G(\bm{x}', \bm{s}_{y_n})$. (Note that $\bm{x}_n$ is a real image at the level $y_n$.)  By this cycle consistency, we have a level-$y$ version of $\bm{x}_n$ and a level-$y_n$ version of $\bm{x}'$. Consequently, we have images at all $L$ levels, even from a single image at a certain level $y$.\par
For generating images at real-valued levels, csGAN introduces an additional loss function, called a {\em order loss}, for $F$:
\begin{equation}
    \label{eq:order loss}
   \mathcal{L}_{\mathrm{order}} = \mathbb{E}_{\bm{z}} \left[ \sum_{y=2}^{L-1}|\bm{s}_{y}(\bm{z}) - \tilde{\bm{s}}_{y}(\bm{z})|\right],  
\end{equation}
where $\tilde{\bm{s}}_{y}(\bm{z}) = (\bm{s}_{y-1}(\bm{z}) + \bm{s}_{y+1}(\bm{z}))/2$. With the order loss, we expect that the style vectors from the same $\bm{z}$ will have a linear property, that is, 
\begin{equation}
   \bm{s}_{y+1}(\bm{z}) - \bm{s}_{y}(\bm{z}) = \bm{s}_{y}(\bm{z}) - \bm{s}_{y-1}(\bm{z}). 
\end{equation}
This linear property will allow us to consider a real-valued severity level $y+\epsilon$, where $\epsilon\in (0,1)$. More specifically, we can derive the style vector for the real-valued level $y+\epsilon$ by the linear interpolation, 
\begin{align}
    \label{eq:style_interpolation}
    \bm{s}_{y+\epsilon} = (1-\epsilon)\bm{s}_{y} + {\epsilon}\bm{s}_{y+1}.
\end{align}
\par
As shown in Fig.~\ref{fig:overview}~(a), at the test phase, we use $G$ to generate an image at a real-valued severity level $y' = y+\epsilon \in [1, L]$.
First, $\{\bm{s}_{1}(\bm{z}), \ldots, \bm{s}_{L}(\bm{z})\}$ is obtained by the mapping network $F$ with a $\bm{z}$. Then, for a certain $\epsilon$, $\bm{s}_{y'}$ is determined by Eq.~(\ref{eq:style_interpolation}). Finally, a level-$y'$ version of an image $\bm{x}$ is generated
by $\bm{x}' = G(\bm{x}, \bm{s}_{y'})$.
\par
\subsection{Learning by Dataset-Disjoint Multi-Objective Optimization}
\label{ssec:cda}
As noted in Section~\ref{sec:intro}, the severity level $y'$ of the generated data is not very reliable. Especially, since we used a simple linear style vector interpolation of Eq.~(\ref{eq:style_interpolation}), we cannot guarantee that the generated data of the level $y'=y+\epsilon$ has exact visual characteristics as the level $y'$. 
In other words, the level $y'$ is not fully reliable as the absolute level. \par
However, $y'$ is still reliable as a {\em relative} level; for a pair of real-valued levels $y'_m$ and $y'_k$ (where $y'_m > y'_k$), the generated images $\bm{x}'_m$ and $\bm{x}'_k$ are expected to show the same relative order in their severity levels, that is, $f(\bm{x}'_m) > f(\bm{x}'_k)$. By training the model $f$ to satisfy this relative condition (instead of training $f$ to satisfy $y_m'\sim f(\bm{x}'_m)$), we can utilize the augmented data  by csGAN in an appropriate manner.\par
Considering the above property of the generated data, we use  {\em dataset-disjoint} multi-objective optimization scheme to train the regression model $f$, as shown in Fig.~\ref{fig:overview}~(b). Assume we have an original dataset $\Omega=\{(\bm{x}_n, y_n)\}$ with manually-annotated discrete severity levels $y_n\in \{1, L\}$ and a generated image dataset $\Omega'=\{(\bm{x}'_m, y'_m)\}$ at various real-valued levels $y'_m\in [1, L]$. Then, the CNN-based regression model $f$ is trained with both datasets $\Omega$ and $\Omega'$ in different usages. Since $y_n$ is reliable as an absolute level, the image $\bm{x}_n$ in $\Omega$ are used to train $f$ to satisfy $y_n\sim f(\bm{x}_{n})$. Here, we use a mean squared error loss $\sum_{n=1}^{N} (f(\bm{x}_{n})-y_{n})^2 /N$. On the other hand, since $y'_m$ is reliable as a relative level, images $\bm{x}'_m$ and $\bm{x}'_k$ in $\Omega'$ with the relative relationship $y_m' > y_k'$ are used to train $f$ to satisfy $f(\bm{x}'_m) > f(\bm{x}'_k)$. Here, we use the loss function of ListNet~\cite{cao2007learning}, which is one of the most popular methods for learning-to-rank. These two loss functions are balanced by a hyperparameter, which is optimized by a validation set.\par
%
\section{Experimental Results}
\label{sec:expr_discuss}
\subsection{Experimental Setup}
\noindent{\bf Dataset:}\ 
To evaluate the proposed method (continuous DA, C-DA in short), we used a dataset of UC endoscopic images collected from the Kyoto Second Red Cross Hospital. The dataset contains 10,265 images from 388 patients. All images are annotated with discrete Mayo scores $y_n\in \{1,2,3,4\}$ by multiple experts and resized to 256 $\times$ 256 pixels. The distribution of Mayo scores is 6,678, 1,995, 1,395, and 197 images for Mayo 0, 1, 2, and 3, respectively. Note that Mayo 0 corresponds to the level $y_n=1$ and Mayo 3 to $y_n=4=L$.
\par
\begin{figure}[t]
    \centering
    \includegraphics[width=0.98\linewidth]{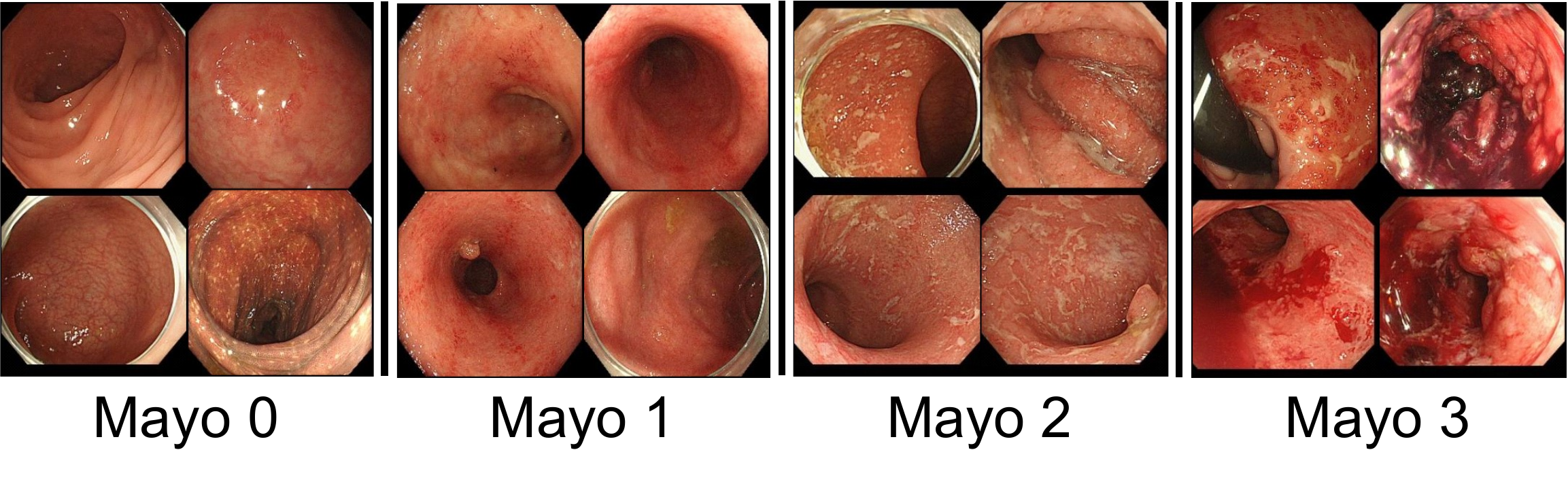}\\[-4mm]
    \caption{Examples of endoscopic images with UC levels.}
    \label{fig:origin_imgs}
    \vspace{-3mm}
\end{figure}
Fig.~\ref{fig:origin_imgs} shows several examples of endoscopic images for each Mayo score. Schroeder et al.\cite{schroeder1987coated} categorized the endoscopic findings of UC as follows: Mayo~0 is a normal or inactive disease, Mayo~1 is a mild disease (erythema, decreased vascular pattern, etc.), Mayo~2 is a moderate disease (marked erythema, erosions, etc.), Mayo~3 is a severe disease (spontaneous bleeding, ulceration, etc.).
\par
We performed five-fold cross-validation. The dataset was divided into training, validation, and test sets at 60, 20, and 20\%, respectively. The splittings were performed by random patient-disjoint sampling, and the class ratios for each set were maintained. Moreover, random oversampling was used to mitigate class imbalance in the training set.
\par

\noindent{\bf Implementation:}\ 
For csGAN, we used the same network structure and hyperparameter values (except that the number of iterations was 50,000) as the official implementation of StarGAN v2 \cite{choi2020stargan}. For the regression model $f$, we used DenseNet \cite{huang2017densely} pretrained on ImageNet \cite{russakovsky2015imagenet} and Adam as the optimizer with the initial learning rate set to $1\times10^{-4}$. The batch size was set to 64. The learning was stopped by the early stopping (no decrease in validation loss for 20 epochs). 

\noindent{\bf Evaluation Metric:}\ 
We quantitatively evaluated the effect of C-DA by the prediction performance of the Mayo score severity classification by $f$. The prediction class (i.e., discrete Mayo level) of the images is determined by quantizing the model outputs into these neighboring discrete levels (e.g.,~1.3~$\rightarrow$~1).
Since the dataset is substantially imbalanced in the number of images in each class, we mainly used the F1 score for the performance evaluation. 
\par
\noindent{\bf Comparative Methods:}\ 
We compared the performance of the proposed DA method (C-DA) with three comparative methods: 1) Baseline, which is conventional regression, 2) Classical DA, which is the baseline with DA by a random combination of horizontal/vertical flipping and rotation, and 3) GAN-based DA, which is used to generate images by the original implementation of a cGAN, called StyleGAN2-ADA \cite{karras2020training}. For 2) and 3), we will show the results with 5,000 generated images per class (i.e., 20,000 in total) because their validation F1 score was saturated even though we used more generated images.
\par
In addition, as an ablation study of C-DA, we evaluated the classification performance of a method that uses the original images $\Omega$ as $\Omega'$ (C-DA w/o GAN). We also performed C-DA under different severity intervals. Specifically, we examined $\epsilon = 1, 0.5,$ and $0.25$ to generate 4, 7, and 13 images from a single $z$, respectively. We used 250 randomly selected $z$s and thus generated 1,000, 1,750, and 3,250 images for each $\epsilon$. Note that the validation F1 scores were almost saturated at 250 $z$s; this means that ours show faster saturations than the above conventional methods, which need 20,000 images ($>3,250$) to saturate.

\subsection{Qualitative Evaluation of Generation Images}

\begin{figure}[t]
    \centering
    \includegraphics[width=\linewidth]{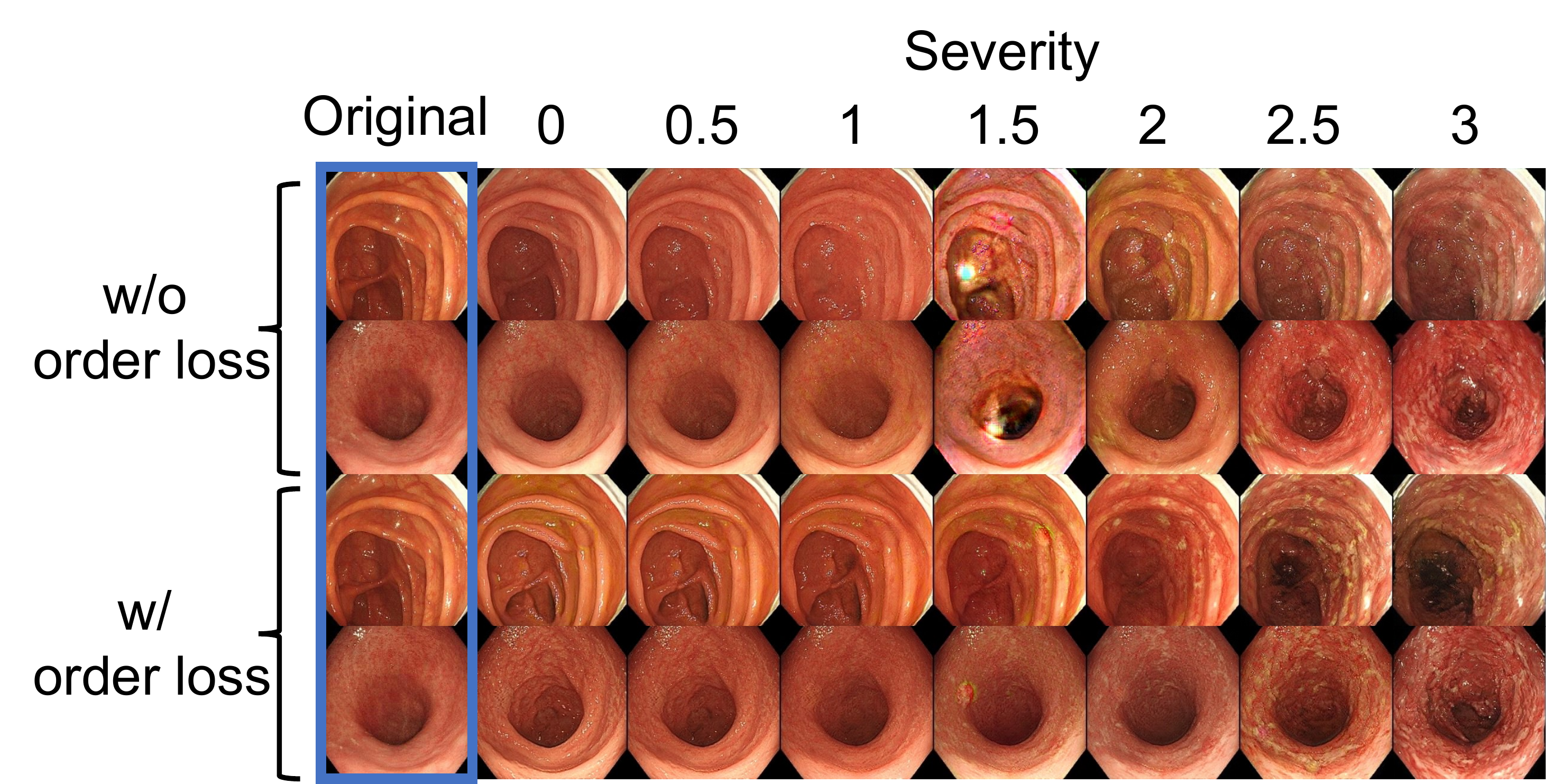}\\[-2mm]
    \caption{Generated images by csGAN at w/o and w/ order loss. 
    The original images at Mayo~0 (in the blue frame) were used to generate the others.}
    \label{fig:gen_imgs}
\bigskip
    \includegraphics[width=\linewidth]{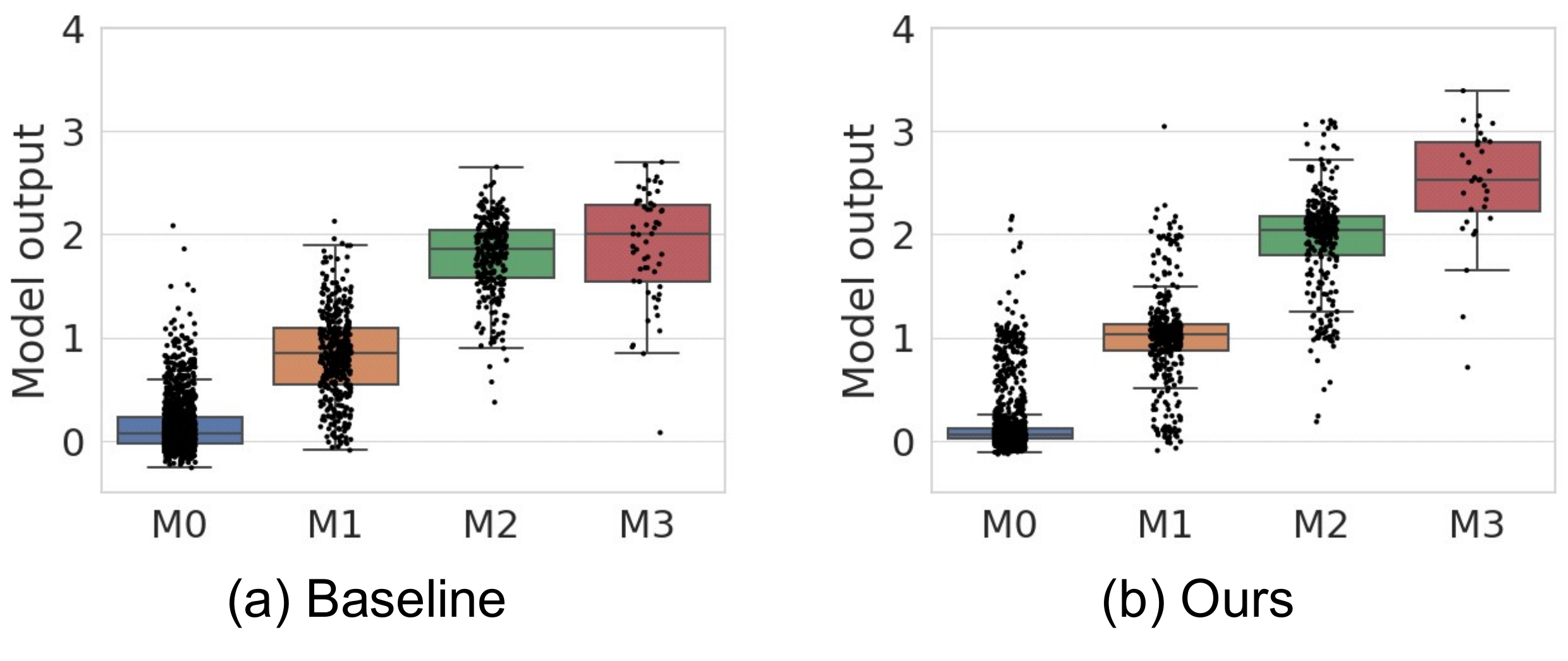}\\[-4mm]
    \caption{Distribution of model outputs $f$ for test images. (a) the regression (Baseline) and (b) the proposed DA (C-DA($\epsilon=0.5$)). ``M0'' stands for Mayo 0.}
    \label{fig:box_plot}
\end{figure}

Fig.~\ref{fig:gen_imgs} shows the generated images with and without order loss by csGAN. Each image was generated with $\epsilon = 0.5$ from an original image at Mayo 0. With the order loss, the severity shifts smoothly between the generated images as the erythema becomes intense, and the semilunar folds gradually disappear as the severity increases. In contrast, without the order loss, the image generated at Mayo 1.5 shows large noises, and the severity is unclear. This observation confirms that the order loss has a stabilization effect of generating images at real-valued levels.

\label{ssec:eval_imgs}

\subsection{Classification Performance}
\label{ssec:eval_cls}

\begin{table}[t]
    \centering
    \caption{Classification performance of each method. ‘*’ denotes a statistically significant difference ($p < 0.05$ in paired t-test) between the baseline and the other methods.}
    \vspace{3.5mm}
    \begin{tabular}{l|ccc}
        \hline
         Method & Precision & Recall & F1-score  \\ \hline
         Regression (Baseline) & {\bf 0.782}$^{\ }$ & 0.631$^{\ }$ & 0.652$^{\ }$\\
         + Classic DA & 0.731$^{\ }$ & 0.657$^{\ }$ & 0.668$^{\ }$ \\
         + GAN-based DA & 0.697$^*$ & 0.651$^{\ }$ & 0.663$^{\ }$ \\ \hline
         C-DA w/o GAN & 0.744$^{\ }$ & 0.629$^{\ }$ & 0.648$^{\ }$ \\ 
         C-DA ($\epsilon=1$) & 0.743$^{\ }$ & 0.624$^{\ }$ & 0.638$^{\ }$ \\
         C-DA ($\epsilon=0.5$) & 0.717$^*$ & {\bf 0.690}$^*$ & {\bf 0.696}$^*$ \\ 
         C-DA ($\epsilon=0.25$) & 0.688$^*$ & 0.672$^*$ & 0.675$^{\ }$ \\ \hline
    \end{tabular}
    \label{tab:performance}
\end{table}
Table~\ref{tab:performance} shows the classification performance of each method. Baseline and conventional DA methods had similar F1 scores, while C-DA ($\epsilon=0.5$) had a higher F1 score than the three comparison methods. These results indicate that the generated images with real-valued severity levels are more effective than conventional DAs. The following facts also confirm this effect. First, the F1 score of C-DA ($\epsilon=0.5$) was higher than that of C-DA w/o GAN. Second, F1 scores of C-DA ($\epsilon=0.5$) were even higher than that of C-DA ($\epsilon=1$). \par

On the other hand, the results also show that image generation at $\epsilon=0.25$ is not very effective. As we noted before, the real-valued levels of the generated images are not completely reliable. Therefore, when $\epsilon$ becomes smaller, the difference between the neighboring levels (e.g., 0.25 and 0.5) becomes unreliable even as the relative levels. 
This fact indicates a limitation in generating images at real-valued levels, and at the same time, it proves the validity of our dataset-disjoint optimization strategy.\par


Fig.~\ref{fig:box_plot} shows box plots of the model output $f$ for test images of each Mayo score. Here, (a) is the regression (Baseline) and (b) the proposed DA (C-DA($\epsilon=0.5$)). The horizontal and vertical axes correspond to the correct Mayo score and the model outputs, respectively. The overall model outputs of C-DA are a narrower interquartile range for each Mayo score than Baseline. Especially, the overlap between the interquartile ranges of Mayo~2 and Mayo~3 is decreased. Consequently, C-DA had better classification performance, even for minor classes with fewer images.

\section{Conclusion}
\label{sec:conslusion}
We proposed a continuous data augmentation (DA) scheme comprising two techniques: continuous severity GAN (csGAN) to generate medical images with real-valued severities and dataset-disjoint multi-objective optimization to utilize the generated images. Through qualitative and quantitative evaluations on an endoscopic ulcerative colitis (UC) image dataset, we confirmed that our DA scheme achieves higher F1 scores by utilizing appropriately generated images. \par
The current limitations of this work are as follows. First, our method is applicable to various tasks with real-valued conditions, and therefore we need to examine our method with other datasets. Second, our UC datasets only have discrete
levels and thus could make our quantitative evaluation in a discrete manner. We will examine different performance evaluations if we find a medical dataset with reliable real-valued annotations. 
\newpage
\section{Compliance with Ethical Standards}
This study was performed in line with the principles of the Declaration of Helsinki. Ethical approval for this study was granted by the Ethics Committee of the Kyoto Second Red Cross Hospital.

\section{Acknowledgments}
This work was supported by JSPS KAKENHI, JP21K18312, and JST SPRING, Grant Number JPMJSP2136.




\bibliographystyle{IEEEbib}
\bibliography{strings,refs}

\end{document}